\documentclass[amsmath,amssymb,twocolumn,amsfonts]{revtex4-2}
\usepackage{graphicx}
\def \beq {\begin{equation}}
\def \eeq {\end{equation}}
\def \ba {\begin{align}}
\def \ea {\end{align}}

\usepackage{bm}
\usepackage{bbm,mathbbol}
\usepackage{braket,bigints}
\usepackage{stmaryrd,mathtools}
\usepackage{csquotes}
\usepackage{upgreek}

\begin{document}
\title{A magnetic falling-sphere viscometer}
\author{C. Patramanis-Thalassinakis\textsuperscript{1}}
\author{P. S. Karavelas\textsuperscript{2}}
\author{I. K. Kominis\textsuperscript{1,2,3}} 
\email{ikominis@qubiom.com}
\affiliation{
$^1$Quantum Biometronics, Heraklion 71409, Greece\\
$^2$Department of Physics, University of Crete, Heraklion 71003, Greece\\
$^3$Institute of Theoretical and Computational Physics, University of Crete, Heraklion 71003, Greece
}
\begin{abstract}
We present a falling-sphere viscometer with a magnetized sphere and fluxgate magnetometers continuously measuring the magnetic field produced at the sensor positions by the falling sphere. With a fluid volume of 15 ml and within a few seconds, we directly measure dynamical viscosities in a range between 200 cP and 3000 cP with a precision of 3\%.
\end{abstract}
\maketitle 
\section{Introduction}
The measurement of the viscosity of (Newtonian) fluids finds applications in several industries, like the pharmaceutical \cite{biomed1,biomed2}, food \cite{food1,food2}, cosmetic \cite{cosmetic1}, and lubricant industry \cite{engine1,engine2}. Based on their operating principles, viscometers can be roughly divided into (i) mechanical, (ii) microfluidic, and (iii) electromagnetic. Early mechanical viscometers still in use are capillary viscometers \cite{cap1,cap2,cap3}, where viscosity is measured by timing the fluid flow through a narrow capillary. Another type of mechanical viscometer measures the torque required to rotate a body (e.g. a disc or a cylinder) inside the fluid \cite{mech1,mech2,mech3,mech4,micromech1,micromech2}. Yet another mechanical viscometer is the falling-sphere viscometer, where the viscosity is found by measuring the terminal velocity of a sphere falling through the fluid under gravity, friction and buoyancy \cite{ball1,ball2,ball3,ball4,ball5,ball6,ball7,ball8,ball9}. Modern microfluidic technology has led to compact devices requiring a small fluid sample volume \cite{microfluid1,microfluid2,microfluid3,microfluid4,microfluid5,microfluid6}. Finally, what we term electromagnetic viscometers are devices using some electromagnetic effect coupled to viscous flow \cite{magn1,magn2,magn3}. For example, a ferrofluid viscometer \cite{magn4} measures the relaxation of a magnetized ferrofluid in the sample under consideration.

We here introduce a falling-sphere viscometer with a \enquote{magnetic twist}. We use a magnetized sphere and fluxgate magnetometers {\it continuously} reading the changing magnetic field produced by the falling sphere at the position of the sensors. By fitting the fluxgate signals to a theoretical form, we can extract the fluid's viscosity with a precision of 3\%. Our viscometer is rather compact (volume occupied by sensors and sample is about 5 cm $\times$ 5 cm $\times$ 10 cm), the measurement time is a few seconds, and the required fluid volume is less than 15 ml. It is worth noting that we directly access the dynamic viscosity of the fluid. In contrast, conventional falling-sphere viscometers measure the sphere's terminal velocity, which depends both on the dynamic viscosity and on the fluid's mass density. Compared to other falling-sphere viscometers, our viscometer is similar to the optical designs using a camera to monitor the sphere's fall \cite{ball3,ball4,ball5,ball6}, in that they both use some physical means (optical versus magnetic) to track the falling ball. While optical viscometers require transparent fluids, such optical designs report a higher accuracy than the one arrived in this work, at an expense of a more elaborate apparatus. One further difference could be cost, however, a direct comparison is not meaningful as the technology and cost of cameras versus fluxgate sensors is changing rapidly. 

In the following section we provide the theoretical description of the experiment presented in Sec. III. In Sec. IV we analyse the measurement results and errors, while in Sec. V we elaborate on several possible sources of measurement uncertainty. In the conclusions of Sec. VI we discuss some possibilities for further developing this methodology. 
\section{Theoretical Description}
Consider a sphere of mass $m$, radius $r$, and mass density $\rho_s=m/{4\over 3}\pi r^3$, moving in a fluid of dynamic viscosity $\eta$. The conventional falling-sphere viscometer measures the sphere's terminal velocity in the fluid, $v_\infty$, under the action of (i) the gravitational force $F_g=mg$, (ii) the Stokes frictional force $F_S=6\pi r\eta v$, and (iii) the buoyant force $F_b={4\over 3}\pi r^3\rho_fg$, where $\rho_f$ is the fluid's mass density. 

Once the falling sphere reaches the terminal velocity under force equilibrium, it will be $F_g=F_S+F_b$, from which equation follows that $\eta=2r^2g(\rho_s-\rho_f)/9v_{\infty}$. The terminal velocity is measured by timing the sphere as it traverses a known distance. Given the fluid's density, the viscosity can be found.

\begin{figure}[t!]
\includegraphics[width=8.5 cm]{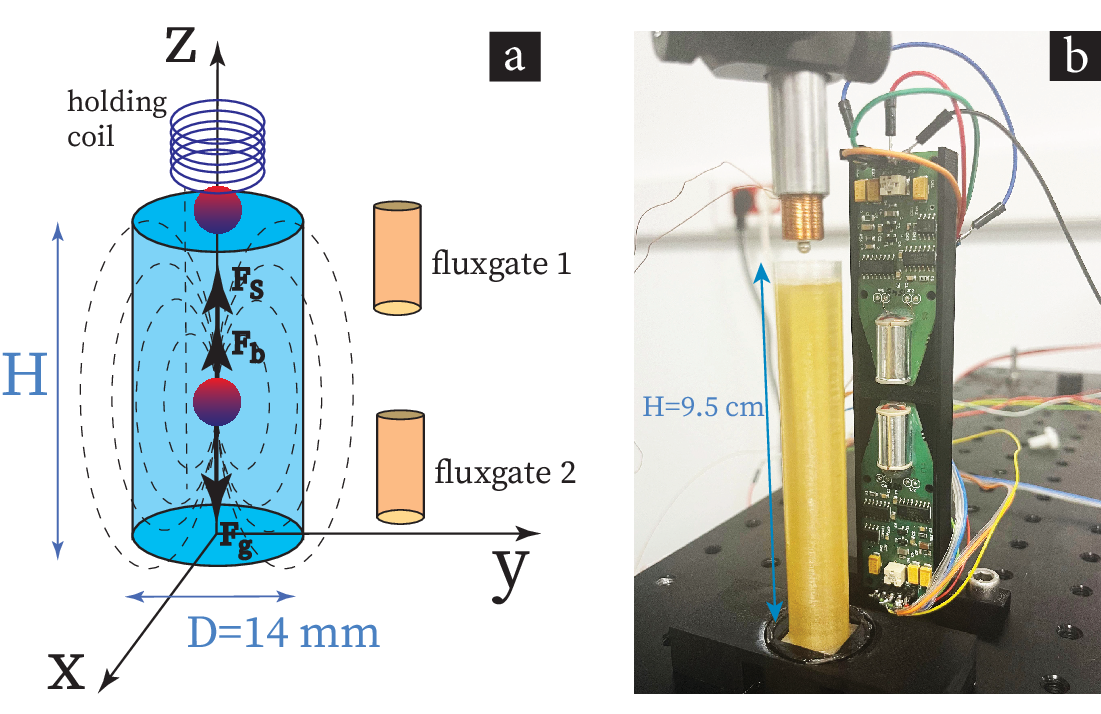}
\caption{(a) Schematic and (b) picture of the magnetic falling-sphere viscometer. A small current-carrying coil holds a magnetic sphere on top of a fluid column. After switching off the current, the sphere falls in the fluid, experiencing the force of gravity, and two opposing forces, the buoyant force, and the frictional Stokes force. Two fluxgate magnetometers are positioned next to the fluid column, and measure the changing magnetic field produced by the falling neodymium sphere. The fluid's dynamic viscosity is extracted by fitting the measured time-dependent magnetic fields to a theoretical form derived herein.}
\label{geom}
\end{figure}

The viscometer presented here does not rely on the measurement of $v_{\infty}$, but on the whole trajectory of the sphere from the top of the fluid column to its bottom, described by the sphere's height as a function of time, $z(t)$. Initially, a neodymium sphere \cite{msds} is held at rest by a current-carrying coil, just above the fluid column's top surface at height $z=H$, as shown in Fig. \ref{geom}a. When the current is switched off at $t=0$, the sphere commences its fall within the fluid. The coordinate system, as shown in Fig. \ref{geom}a, has the coordinate center at the bottom and center of the cylindrical fluid column.

The height of the sphere can be found by solving the equation of motion $m\ddot{z}=-mg-6\pi r\eta \dot{z}+F_b$, with $F_b$ as given before. The initial conditions are $z(0)=H$ and $\dot{z}(0)=0$. Defining the time constant $\tau\equiv 2\rho_s r^2/9\eta$, it follows that 
\beq
z(t)=H+g\big(1-{\rho_f\over \rho_s}\big)\tau^2\Big(1-{t\over\tau}-e^{-t/\tau}\Big)\label{zt}
\eeq
The time constant $\tau$ quantifies the time it takes for the sphere to reach terminal velocity, i.e. when the exponential term in Eq. \eqref{zt} has become negligible. We stress that our measurement does not rely on the sphere reaching terminal velocity, i.e. when analyzing data with the theoretical model we use the exact expression of Eq. \eqref{zt}.  

From Eq. \eqref{zt} it is seen why we can directly access the dynamic viscosity $\eta$, or equivalently the time constant $\tau$, from the measured signal. This is because the sphere's trajectory $z(t)$, which underlies the magnetometers' signal as detailed next, has a dependence on the parameter $\tau$ different from the dependence on the fluid's density $\rho_f$. By inspecting Eq. \eqref{zt} one might think that this different dependence is due to the exponential term $e^{-t/\tau}$. But actually, this different dependence remains even after reaching terminal velocity. Indeed, when the exponential $e^{-t/\tau}$ in Eq. \eqref{zt} becomes negligible, the sphere's height as a function of time will read $z_{\infty}(t)=H+g(1-\rho_f/\rho_s)\tau^2(1-t/\tau)$. In this expression $\rho_f$ and $\tau$ are still decoupled when using many data points at different times $t$. In contrast, conventional falling-sphere viscometers measure one number, the terminal velocity. By taking the time derivative of $z_{\infty}(t)$, it is seen again that such viscometers are sensitive to $|\dot{z}_{\infty}|=g(1-{\rho_f\over \rho_s}\big)\tau$. Thus, $\rho_f$ and $\tau$ cannot be individually found from the measurement of this product.
\subsection{Measurable Viscosity Range}
The SI unit of viscosity is 1 ${\rm Ns/m^2}=1000~{\rm cP}$. For example, the viscosity of engine oil at room temperature is about 500 cP. Given the sphere's density $\rho_s\approx 7.47~{\rm g/cm^3}$, it follows that the corresponding value of the parameter $\tau$ is 4 ms. The sphere's density was estimated from the mass and radius data given by the manufacturer for much larger spheres of the same material, in order to minimize the relative error of the estimate \cite{neod}.

To find the range of values of $\tau$ measurable with our methodology we first note that, as is evident from Eq. \eqref{zt} by expanding the exponential term to second order, for $t\ll\tau$ the height of the sphere $z(t)$ ceases to depend on $\tau$. Thus a small viscosity (large $\tau$) is not measurable using a too short trajectory, since the sphere will practically undergo free fall at early times. For example, the time to reach the bottom of our 9.5 cm cylinder by free fall is about 0.15 s, hence this would be an approximate upper limit for the measurability of the parameter $\tau$ with such a device, translating into a lower limit for the viscosity of $\eta\approx 10-20~{\rm cP}$ for typical fluid densities. The upper limit of the measurable viscosity can in principle be arbitrarily high, as long as the sphere does fall through the fluid.
\subsection{Magnetometer Signals}
\begin{figure}[th!]
\includegraphics[width=7cm]{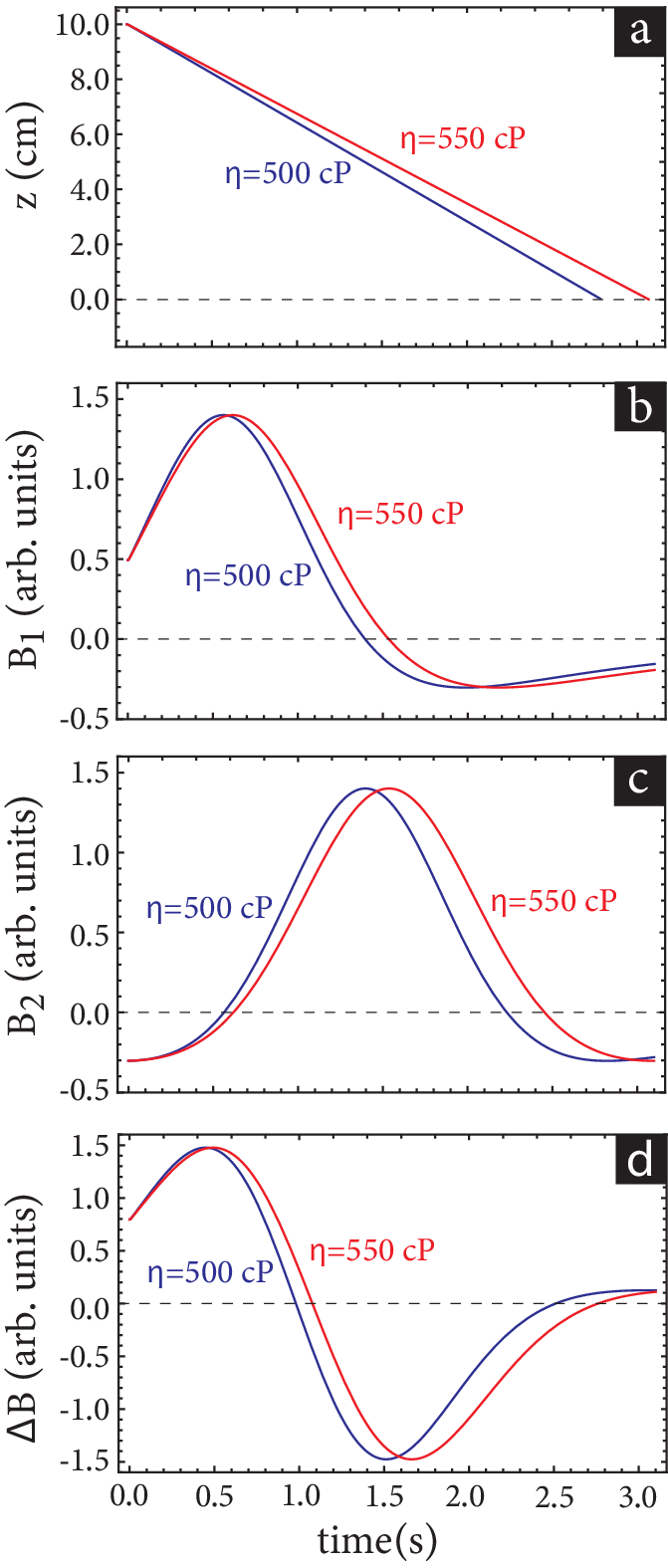}
\caption{Calculated examples of falling sphere trajectory and magnetometer signals for two values of the viscosity, $\eta=500~{\rm cP}$ (blue curves), and  $\eta=550~{\rm cP}$ (red curves). The positions of the two sensors were $a=4 ~{\rm cm}$, $b=0$, $c_1=8~{\rm cm}$ for the upper sensor, and $c_2=5~{\rm cm}$ for the lower sensor. (a) Height of the ball as a function of time. (b) Signal $B_1$ of the upper sensor. (c) Signal $B_2$ of the lower sensor. (d) Difference signal $\Delta B=B_1-B_2$. Parameter values for this calculation were: sphere radius and density $r=1.46~{\rm mm}$ and $\rho_s=7.47~{\rm g/cm^3}$, respectively, fluid density $\rho_f=0.8~~{\rm g/cm^3}$, fluid column diameter $D=14~{\rm mm}$, fluid column height $H=10~{\rm cm}$.}
\label{plots}
\end{figure}
The magnetic field produced by a magnetic dipole of moment $\mathbf{m}$ at the position vector $\mathbf{r}$ with respect to the dipole is $\mathbf{B}[\mathbf{r}]={\mu_0\over {4\pi}}({{3(\mathbf{m}\cdot\mathbf{r})\mathbf{r}}\over {r^5}}-{\mathbf{m}\over r^3})$, where $r=|\mathbf{r}|$. As shown in Fig. \ref{geom}, we use two fluxgate sensors adjacent to the fluid, with their sensitive axes being along the z axis, the sphere's trajectory. For the moment we consider point sensors, and later we will take into account the finite sensing volume. Let the position of the $j$-th fluxgate sensor be denoted by the position vector $(a,b,c_j)$, where $j=1,2$. That is, we consider the two point sensors to define a line parallel to the z-axis. Then, the position of the $j$-th sensor with respect to the falling sphere is $\mathbf{r}_j=a\mathbf{\hat{x}}+b\mathbf{\hat{y}}+\big(c_j-z(t)\big)\mathbf{\hat{z}}$. Thus, the signal of the $j$-th sensor will be $B_j(t)=\mathbf{\hat z}\cdot\mathbf{B}\big[\mathbf{r}_j\big]$. 

At time $t=0$ the magnetization of the sphere is aligned with the axis of the current-carrying coil, the $z$-axis. Setting $\mathbf{m}=m\mathbf{\hat z}$ and $B_0=\mu_0 m/4\pi$ we find
\beq
B_j(t)=B_0{1\over {(a^2+b^2)^{3/2}}}{{2f_j^2(t)-1}\over {\big[1+f_j^2(t)\big]^{5/2}}}+b_0,\label{Bj}
\eeq
where 
\beq
f_j(t)={{c_j-z(t)}\over\sqrt{a^2+b^2}},\label{fj}
\eeq
and $b_0$ a background magnetic field common to both sensors. By measuring the difference $\Delta B=B_1(t)-B_2(t)$, the background field drops out. This helps suppress common magnetic fields, in particular ac magnetic fields from nearby 50 Hz power lines. In summary, the viscosity $\eta$ hides in the parameter $\tau$ entering the sphere's height $z(t)$ given by Eq. \eqref{zt}, which in turn enters the measured magnetic fields $B_j(t)$ through Eqs. \eqref{fj} and \eqref{Bj}.

In Fig. \ref{plots} we present example plots for the sphere's trajectory $z(t)$ (Fig. \ref{plots}a), the signals $B_1(t)$ (Fig. \ref{plots}b) and $B_2(t)$ (Fig. \ref{plots}c), and the difference $\Delta B=B_1-B_2$ (Fig. \ref{plots}d), for two values of the viscosity. For generating these plots, we considered two corrections of the simplified description outlined previously. 

First, the correction due to the finite volume of the fluid column, the so-called edge effect. This has been discussed in detail in \cite{ball3,ball4,ball8}, and we here follow the treatment presented therein. In particular, the measured viscosity overestimates the true viscosity, because the walls of the fluid's container effectively push the sphere upwards. This is quantified by a correction factor $K_{\rm edge}$, which for small Reynolds numbers pertinent to our measurements (${\rm Re}\lessapprox 2$) is given by \cite{Sayre}
\beq
K_{\rm edge}={{1+n_5x^5}\over {1+d_1x+d_3x^3+d_5x^5+d_6x^6}},
\eeq
where $x=2r/D$, $n_5=-0.75857$, $d_1=-2.1050$, $d_3=2.0865$, $d_5=-1.7068$, and $d_6=0.72603$. For our case, with $D=14~{\rm mm}$ being the diameter of the cylindrical fluid column, and $r=1.46~{\rm mm}$ being the sphere's radius, it is $K_{\rm edge}=1.726$.
\begin{figure*}[ht!]
\includegraphics[width=17.5 cm]{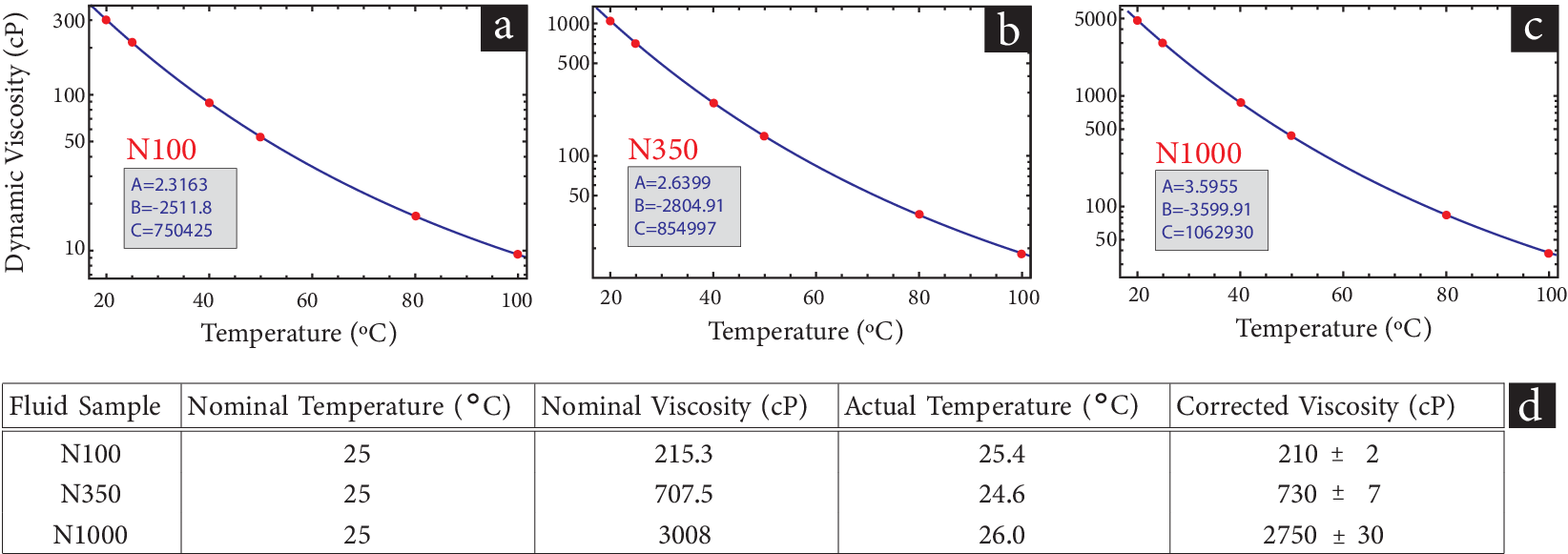}
\caption{Temperature dependence of the viscosity, $\eta(T)$, as given by the manufacturer \cite{rheotek} at 6 different temperatures for three reference standard oils: (a) N100, (b) N350, and (c) N1000. Solid line is a fit to the functional form $\log\eta(T)=A+B/T+C/T^2$, with the respective fit parameters $A$, $B$ and $C$ shown in the insets. In the formula $T$ is the absolute temperature. We use the fit to correct for the standard viscosity, since our measurements were performed around $25~^\circ{\rm C}$, but not exactly at $25~^\circ{\rm C}$, which is the second temperature data point provided by the manufacturer. The fit parameter errors are negligible. (d) Table shows the nominal viscosity reference values at $25~^\circ{\rm C}$, and the actual values calculated from the fit at the actual temperature of our measurements. The errors quoted in the values of the corrected viscosities in the last column of the table derive from an uncertainty of $0.1~^\circ{\rm C}$ in the actual temperature shown in the third column of the table.}
\label{temp}
\end{figure*}

The second correction is due to the fact that the fluxgate sensors are not point sensors, but have a finite volume of a strip geometry with length 2.2 cm, width 1.5 mm and thickness 0.025 mm. To simulate the sensor signal, we thus integrate the magnetic field produced by the sphere in the finite volume of the sensor. The theoretical fits to the data presented next include both aforementioned corrections.
\section{Experiment}
To test the magnetic viscometer we used three viscosity standards \cite{rheotek}, which were oils of known viscosity ranging from about 200 cP to 3000 cP. The viscosity reference is given by the manufacturer at 6 different temperatures. We used the reference values at $25~^\circ{\rm C}$, but our measurement was not performed exactly at $25~^\circ{\rm C}$. Thus we fitted the temperature dependence of each standard, and from the fits we found the standards' viscosity at the actual measurement temperature. In Figs. \ref{temp}a-c we show the temperature dependence of the three viscosity standards, together with the theoretical fits to the functional form $\log\eta(T)=A+B/T+C/T^2$, suggested in \cite{temp}. In the table of Fig. \ref{temp}d we present the nominal values of the viscosity standards at $25~^\circ{\rm C}$, along with the corrected values at the actual temperature of our measurement and the corresponding error. As the manufacturer does not quote any errors in the reported standard values, we used as error source a $\pm 0.1~^\circ{\rm C}$ uncertainty in the temperature of the fluid, leading to an uncertainty in the reference viscosity value around 1\%. 

\begin{figure}[th!]
\includegraphics[width=7. cm]{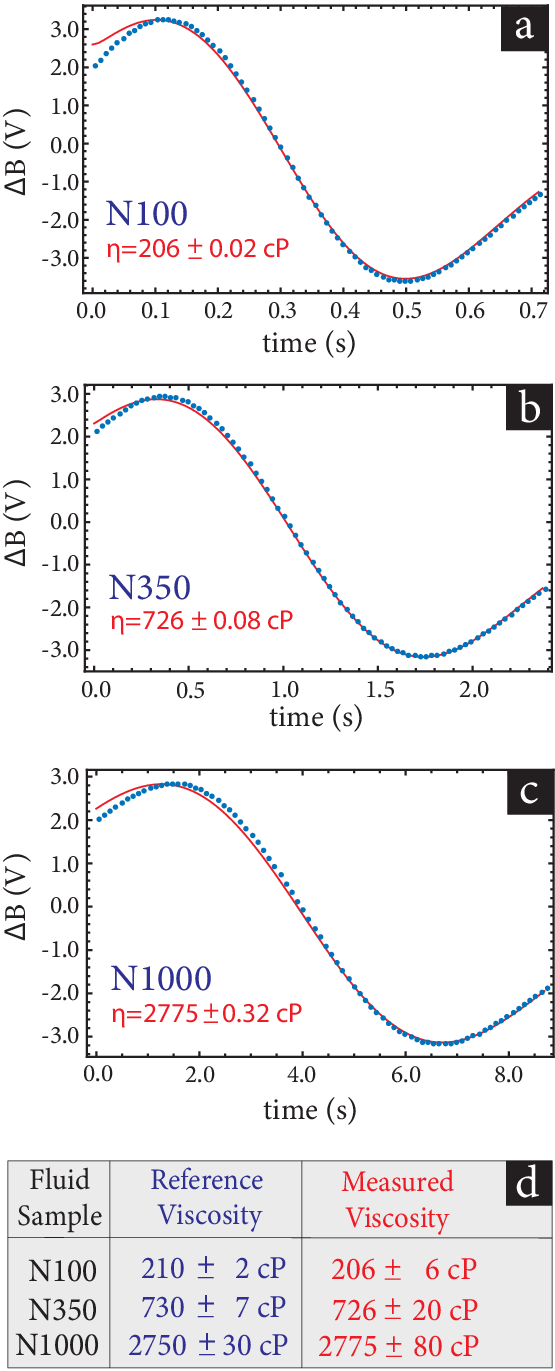}
\caption{(a-c) Measured data and theoretical fits for three viscosity reference standards. The measurement $\Delta B$ is the difference of the two fluxgate output signals (in Volt) as a function of time. In each plot we show the result of the Levenberg-Marquardt fitting algorithm for the viscosity of the tested fluid and the fit error. (d) All three viscosity values resulting from the fits, together with the corresponding standard values obtained as shown in Fig. 3. Final measurement error is 3\%, as follows by repeating the measurements with the same sample.}
\label{data}
\end{figure}

In Figs. \ref{data}a-c we present the actual measurements for the three standard oils, together with the fits to the theoretical form of Eq. \eqref{Bj}, including the corrections mentioned in Sec. II. The output voltages of the two fluxgate sensors where digitized with a National Instruments DAQ card at an acquisition rate of 1 kHz. The presented measurements are the differences, $\Delta B$, of the signals recorded by the two fluxgate sensors. An average in time was then performed so that in all cases the final measured trace for $\Delta B$ has 80 data points. The duration of the measurement was defined by the time when the lower sensor reads a maximum value, increased by 50\%. This way we omit the data points originating from the sphere's trajectory close to the bottom of the container, in order to not have to include additional corrections due to the finite length of the trajectory \cite{ball3,ball4,ball8}.

The fits were obtained with the Levenberg-Marquardt algorithm \cite{Strutz}, using as fitting parameters the viscosity, the fluid's density, an amplitude scaling the overall signal, and an additive offset. The positions of the two sensors relative to the fluid column, and the initial height of the sphere were measured and kept constant. In particular, $a=3.3~{\rm cm}$ and $b=4~{\rm mm}$ define the lateral positions of the center of the sensors, while $c_1=8.4~{\rm cm}$ and $c_2=5.4~{\rm cm}$ was the height of the upper and lower sensor, respectively. The initial sphere's height was $H=9.5~{\rm cm}$. The fitted signal amplitudes were 1.28 V, 1.14 V and 1.12 V, corresponding to the samples N100, N350 and N1000. If the sphere has the exact same trajectory with respect to the sensors in every measurement, these three amplitudes should be the same. The amplitude for N100 is 13\% larger than the rest, which can be explained by a slightly shifted placement of the current-carrying coil with the sphere (in every trial we first attach the sphere at the tip of the current carrying coil, and then lower the coil to visually position the sphere just on top of the fluid surface).

As seen in Figs. \ref{data}a-c, there is excellent agreement of the theoretical fits with the measured data. The slight discrepancy between data and fits observed in the beginning phase of the signals is conceivably due an interplay of effects not taken into account in our theoretical model and related to the splash of the sphere on the liquid surface \cite{wetting}, to surface tension, and wetting.
\section{Results and Error Analysis}
In each of the Figs. \ref{data}a-c we also display the result for the fit parameter $\eta$, together with the error resulting from the fit. This is calculated by \cite{Strutz} $\delta(\Delta B)/\sum_{j=1}^{80}(\partial f(t_j)/\partial\eta-(\Delta B)_j)^2$, where $\delta(\Delta B)\approx 5~{\rm mV}$ is the measured noise in $\Delta B$, $\Delta B_j$ are the measured values of $\Delta B$ at time $t_j$, and $\partial f(t_j)/\partial\eta$ the sensitivity of the theoretical form to $\eta$ at time $t_j$.  Incidentally, the quoted intrinsic noise of our fluxgate sensors is $20~{\rm pT/\sqrt{Hz}}$ at 1 Hz, which within the 1kHz bandwidth translates to about $\delta(\Delta B)=100~{\rm \mu V}$ noise. Our noise level of 5 mV is mostly due to the sensors operating in the unshielded environment of the lab, without any filters to reduce low-frequency noise. In any case, as will be shown next, the quoted fit-parameter errors stemming from the noise in $\Delta B$ are negligible. Nevertheless, this points to the possibility to obtain, in principle, even lower uncertainties in the estimate of the viscosity, which would take advantage of the intrinsic noise level of the sensors.  

The fit errors shown in Figs. \ref{data}a-c underestimate the precision of our measurements. This is seen by repeating the measurement with the same sample (10 repetitions), in which case we get a relative standard deviation in the viscosity estimates around 3\%. This is the final quoted measurement error in the table of Fig. \ref{data}d, which shows how the measured viscosities compare with the corresponding standard values, demonstrating a very good agreement given the simplicity of our setup. Sources of the 3\% variability could be short-term temperature drifts, or small rotations of the sphere due to small density inhomogeneities (small bubbles) of the fluid. 

Finally, as noted in the introduction, one practical advantage of our methodology is that it allows to directly access the dynamic viscosity $\eta$, in contrast to conventional falling-sphere viscometers requiring knowledge of the fluid's density in order to extract the dynamic viscosity from the measurement of the terminal velocity. While we do leave the fluid density $\rho_f$ as a fitting parameter in the fitting algorithm, the data reported herein do not provide for a precise measurement of $\rho_f$. This is seen qualitatively by the fact that $\rho_f$ enters Eq. \eqref{zt} through the expression $1-\rho_f/\rho_s$, and the ratio $\rho_f/\rho_s$ is about 0.1, hence the fluid density only mildly affects the sphere's trajectory. 

Quantitatively, it becomes evident that the $\chi^2$ dependence on $\rho_f$, where $\chi^2=\sum_{j=1}^{80}(f(t_j)-(\Delta B)_j)^2$, has a very shallow minimum. In particular, $\chi^2$ is about two orders of magnitude less sensitive on $\rho_f$ than it is on $\eta$. The result is that the fitted values for the fluid's mass density indeed follow the trend of the numbers reported by the manufacturer for the three standard fluids (N100: $0.874~{\rm g/cm^3}$ , N350: $0.891~{\rm g/cm^3}$, N1000: $0.921~{\rm g/cm^3}$), but are about 5\%-20\% off. Moreover, the exact discrepancy depends on the particular parameter update step chosen in the fitting algorithm. Such variability in the density estimate translates into viscosity estimate changes within the 3\% error quoted above.
\section{Discussion of measurement uncertainties}
Since this viscometer is based on measuring the magnetic field produced by the magnetized sphere, it should be operated away from ambient magnetic-field sources, in particular strong magnets that might saturate the sensors or affect the sphere. If this is not an issue, it could be time-dependent magnetic fields different at the two sensors that could cause extra noise, since the difference signal removes common mode noise, while a signal difference constant in time is taken care of by the fitted background of the difference signal $\Delta B$. 

Regarding possible forces on the magnetic sphere, the ferromagnetic core of the fluxgate sensors themselves produces a magnetic field, thus we placed the sensors at a horizontal distance of 3.3 cm from the fluid sample. We measured the magnetic gradient produced by the sensors at the position of the fluid, and it was found to be around 1 mG/cm. Taking into account the remanence of the sphere (1.3 T), we estimate the sphere's magnetic moment and find that the force on the sphere due to this gradient is four orders of magnitude smaller than the sphere's weight. Hence the sensors themselves do not affect the sphere. In any case, if this viscometer is required to operate close to strong laboratory magnets, it should be enclosed in a magnetic shield. 

Regarding possible rotation of the sphere upon release from the current-carrying coil, if there was such a rotation the theoretical model would not be able to fit the data, since the theoretical morel assumes constant magentization of the sphere along the z-axis (the sphere's trajectory). Nevertheless, we also used a magnetized cylinder and visually inspected the fall, which did not exhibit any noticeable rotation upon release.

Another concern, due to the sensitive temperature dependence of viscosity, could be heating of the fluid sample by the frictional Stokes force. With an-order-of-magnitude calculation it is seen that such an effect should be negligible. Indeed, for a fluid specific heat on the order of $1~{\rm J/g/^{\circ}K}$ and setting the work done by the sphere's weight (equal to the to opposing forces when in equilibrium) equal to the heat transferred to the fluid, we find a temperature change on the order of ${\rm \upmu K}$.

Yet another concern could be the fluid's density fluctuations, possibly causing random rotations of the magnetized sphere, and secondly causing small random deviations from the trajectory of Eq. \eqref{zt} due to a random change in the buoyant force. However, thermodynamic density fluctuations resulting from particle number fluctuations, on the order of $1/\sqrt{N}$ where $N$ the number of fluid particles \cite{density} in the macroscopic volume occupied by the sphere are negligible, since $N\approx 10^{20}$. On the other hand, there could be density fluctuations due to more rudimentary issues like tiny bubbles in the fluid. These, however, are much harder to systematically quantify. 

Finally, we checked the sphericity of the spheres and found the non-sphericity to be at the level of 0.5\%, which translates into 1\% uncertainty in the parameter $\tau$, due to the $r^2$-dependence of $\tau$. This error, along with other uncertainties considered in \cite{ball3,ball4} is negligible with respect to the precision of 3\%, which also reflects the accuracy of this measurement. Summarizing, in this work our aim is not to compete with previous realizations of falling-sphere viscometers in terms of precision/accuracy, but to introduce a new kind of falling-sphere viscometer, the precision and accuracy of which we hope to improve in future refinements of the method. 
\section{Conclusions}
We have presented a simple falling-sphere viscometer using a magnetic sphere and two fluxgate sensors continuously monitoring the sphere's fall within the test fluid. The viscometer's precision could be further improved by modifying the design details of this methodology, in particular the temperature stability. The fluid volume used in this work is 15 mL, and it can be further reduced by using a smaller diameter sphere and a smaller fluid container. The method can also work at higher temperatures, at least up to $80~^\circ{\rm C}$ quoted by the magnetized sphere manufacturer, which is a fraction of neodymium's Curie point. One could even conceive a significant miniaturization of this technique towards measuring ultra-low fluid sample volumes by using different kinds of magnetometers, like diamond sensors \cite{diamond1, diamond2}, or miniaturized atomic magnetometers \cite{Kitching1,Kitching2}.

We acknowledge the co-financing of this research by the European Union and Greek national funds through the Operational Program Crete 2020-2024, under the call "Partnerships of Companies with Institutions for Research and Transfer of Knowledge in the Thematic Priorities of RIS3Crete", with the project title "Analyzing urban dynamics through monitoring the city magnetic environment" (Project KPHP1-No. 0029067).


\begin{thebibliography}{0}
\bibitem{msds}
The user of this methodology should consult the MSDS of the fluids under test in order to gauge their compatibility with neodymium or any other material the falling sphere is made of.
\end{thebibliography}
\end{document}